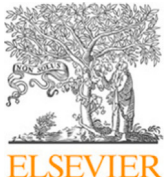
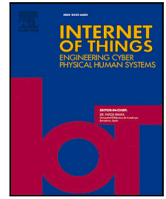

Research article

# An IoT system for a smart campus: Challenges and solutions illustrated over several real-world use cases

Tomás Domínguez-Bolaño [*], Valentín Barral, Carlos J. Escudero, José A. García-Naya

*CITIC Research Center & Department of Computer Engineering, University of A Coruña, 15071, A Coruña, Spain*

A B S T R A C T

This article discusses the development of an IoT system for monitoring and controlling various devices and systems from different vendors. The authors considered key challenges in IoT projects, such as interoperability and integration, scalability, and data storage, processing, and visualization, during the design and deployment phases. In addition to these general challenges, the authors also delve into the specific integration challenges they encountered. Various devices and systems were integrated into the system and five real-world scenarios in a university campus environment are used to illustrate the challenges encountered. The scenarios involve monitoring various aspects of a university campus environment, including air quality, environmental parameters, energy efficiency, solar photovoltaic energy, and energy consumption. The authors analyzed data and CPU usage to ensure that the system could handle the large amount of data generated by the devices. The platform developed uses open source projects such as Home Assistant, InfluxDB, Grafana, and Node-RED. All developments have been published as open source in public repositories. In conclusion, this work highlights the potential and feasibility of IoT systems in various real-world applications, the importance of considering key challenges in IoT projects during the design and deployment phases, and the specific integration challenges that may be encountered.

## 1. Introduction

Companies and organizations have traditionally used various systems to monitor and respond to different needs such as heating, ventilation, and air conditioning (HVAC), leak detection, presence, and luminosity. In many cases, these systems were isolated, implemented independently of one another by different departments within the organization, such as occupational safety and health, infrastructure, information technology (IT), and security.

The Internet of things (IoT) has made this situation even more complex, as more and more devices are being deployed that produce data that needs to be monitored and managed. In this regard, we may find that today many of the IoT device vendors offer proprietary and closed cloud-based solutions. However, these would contribute to the dispersion and heterogeneity of the monitoring systems, and in addition, some other limitations may soon appear, such as:

- Maintenance of a cloud platform with the typical problems exhibited by any platform: access control management, potential software and hardware vulnerabilities, and periodic payment of licenses.
- Vendor lock-in. We may be locked into a vendor's solution without the ability to integrate sensors from other vendors.

---

[*] Corresponding author.
  *E-mail addresses:* tomas.bolano@udc.es (T. Domínguez-Bolaño), valentin.barral@udc.es (V. Barral), escudero@udc.es (C.J. Escudero), jagarcia@udc.es (J.A. García-Naya).

- Limited access to the data for analysis. In a vendor-controlled platform, functionalities may be predefined, with little or no control over them. It is desirable for the vendor to provide an application programming interface (API) that allows external access to the data, but sometimes it does not exist or the data provided does not follow a standard.
- Difficulties in cross-referencing data from different sources. As an example that arises during this work, consider one data source of environmental parameters: temperature, humidity and air quality levels obtained from multiple spaces and rooms in an organization; and another data source of estimates of the occupancy of those spaces, based, for example, on the number of devices connected to Wi-Fi access points. It is desirable to be able to cross-reference both data sources to analyze the effect of the occupancy of each space on its respective environmental parameters, which will allow for improved performance of HVAC systems with the goal of achieving an optimal balance between health and energy efficiency. With proprietary and closed systems, these data sources may be isolated and therefore difficult to analyze together.

Therefore, while these solutions may be appropriate in simple cases, they can be problematic for larger environments, where a significant number of IoT devices from different vendors and using different standards may need to be integrated. To address this complexity, we need, on the one hand, open and interoperable IoT systems that allow devices and systems from different vendors to work together seamlessly, and on the other hand, we need a centralized platform to store the data and manage the various monitoring and action systems. In this way, the so-called smart environments can be build, where sensors, actuators, and the analysis of their data enable intelligent and responsive environments that are able to adapt to the needs and preferences of the users, as well as to the changing environmental conditions. Such environments can be found in a variety of settings, including homes, offices, universities, and cities. Overall, smart environments, making use of open and interoperable IoT technologies, are transforming the way that businesses and organizations operate and compete. By leveraging the power of these systems, organizations can gain greater insight, control, and efficiency in their operations.

This article presents a real case study of an IoT platform that arose at the University of A Coruña (UDC), in Spain, due to the situation caused by the pandemic triggered by the SARS-CoV-2 virus that appeared at the end of 2019. This situation demonstrated the need for an IoT platform capable of responding quickly to new needs. It was in response to the rapid propagation of the SARS-CoV-2 virus that governments and institutions sought mechanisms to contain its spread and minimize the impact of the disease it causes (COVID-19). In particular, one of the most important aspects has been the control of the air quality since it has been demonstrated that the transmission of these viruses is through aerosols in the air [1–3]. The most direct way to control air quality and aerosols is to monitor $CO_2$ concentrations [4,5]. Thus, the need arose in our institution for an IoT platform to collect real-time information on $CO_2$ levels in the different spaces (i.e., classrooms, study rooms, auditoriums, common areas, and workspaces).

To address the above situation, the university's occupational safety and health department had acquired a large number of $CO_2$ sensors from a specific manufacturer with a license to use them with their proprietary cloud system. However, this not only had several limitations as explained above but it also exacerbated the dispersion and heterogeneity of monitoring systems in our institution. Therefore, we decided to deploy an IoT platform that would allow us to monitor and control not only the $CO_2$ with the sensors that we already had, but also other devices and systems from different manufacturers. With this platform we were able to implement real-time $CO_2$ concentration monitoring in several of the university's buildings. In particular, the system was considered for real-time monitoring of the 2021 university admission exams (the measurement data for these exams were published in [6]). The data collected allowed for safer exams from the point of view of the SARS-CoV-2 pandemic, while at the same time the data were used for several statistical studies [7,8].

After that, the platform proved to be very useful for several other monitoring tasks at the university. In particular, the most prominent are:

- Monitoring environmental parameters at the Center for Information and Communications Technology Research (CITIC), a research center[1] of the UDC, Spain. Sensors were installed throughout the building of the CITIC to monitor various environmental parameters such as temperature, humidity, and $CO_2$.
- Monitoring of systems deployed by the architecture, urban planning and equipment department of our university. This department is in charge of several important tasks such as the planning and equipment of the buildings, maintenance of several installations, and energy management. Thus, using the aforementioned platform, we implemented real-time monitoring of different parameters in several systems, such as the energy production of several photovoltaic systems, the energy consumption in different areas of the university, and other environmental parameters such as the radon gas concentration in several rooms.

Deploying an IoT system can be a complex process and many challenges and issues can arise. The contributions of the various sections of this paper can be summarized in the following points:

1. In Section 2 we review the key IoT challenges and we detail how each of them influenced the design of our IoT system. These key challenges were analyzed previously in the literature, but only from a theoretical point of view. In contrast, in this paper we explain how they have been considered in a real case.
2. In Section 3 we explain the specific challenges we encountered when integrating IoT devices with our IoT system. To the authors' knowledge, these specific integration challenges have not been previously described in the literature.
3. In Section 4 we detail the integration of several IoT devices with our IoT system. This provides several concrete real-world application examples that illustrate the various integration challenges presented in Section 3.

---

[1] https://citic.udc.es/en/

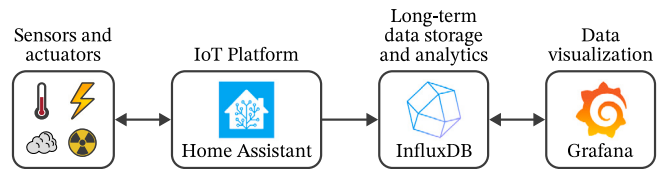

**Fig. 1.** System architecture.

4. One of the problems with IoT systems is that the sheer volume of data generated by the growing number of devices can quickly become overwhelming, so it is essential to ensure that the system resources are scaled appropriately. Therefore, in Section 5 we provide CPU and data usage statistics of our system.
5. The main purpose of the IoT system shown in this paper is to monitor the data provided by the sensors. Therefore, for completeness, in Section 6 we show some examples of monitoring that we have performed.
6. Finally, several software developments were carried out for the integrations shown in Section 4. Therefore, in Section 7 we describe these developments and provide the links to the public repositories where they are available.

## 2. Deployment and development of our IoT system taking into account key challenges

As mentioned in the introduction, we wanted to deploy an IoT system that would allow us to monitor and control various heterogeneous devices and systems from different vendors. Fig. 1 shows the architecture of the deployed system. In this architecture we show the sensors and actuators as an abstract block, but note that our work so far has only considered the monitoring of different parameters from various environments and systems (i.e., temperature, electrical power, $CO_2$ concentration, radiation, etc.), hence in this manuscript we will only consider the sensor part.

IoT projects are usually complex and face several crucial challenges that need to be considered. Many of these challenges have already been analyzed in the literature [9–12], but only from a theoretical point of view. When deploying IoT systems using enterprise cloud platforms, the cloud provider already takes care of many of these challenges, but in this work we deployed the IoT systems on-premises. Therefore, below we list some of the most important of these challenges, and for each one we describe how we considered it and how it influenced the design of our system:

1. **Interoperability and integration**: IoT devices from different manufacturers and using different network and data communication standards should be able to work together seamlessly. This is essential in smart environments where many types of devices may need to be connected and coordinated in order to monitor and control different aspects of facilities and infrastructures. For this task, one of the most common solutions in practice is the use of a centralized middleware software to manage IoT devices and data from a single point [10]. This middleware is what we will call an IoT platform, and it is the approach that we followed for deploying the IoT system presented in this work.
   At this point, we carefully evaluated the different alternatives available [13] and decided to use the Home Assistant IoT platform [14]. This decision was motivated by its extensive development community (which already provides an impressive amount of integrations with different types of sensors, manufacturers, and external platforms), the ease of developing new integrations and extending their functionalities, and previous success stories such as those in [15–17].
2. **Cost**: Deploying and maintaining an IoT system can be expensive, especially if it involves a large number of devices and/or requires specialized infrastructure. It is important to carefully consider the total cost of ownership of the devices and infrastructure. In this regard, we decided to deploy the system on premises using free and open-source software components, as the CITIC research center provided a modern data center that we could use for the deployments.
3. **Data storage, processing and visualization**: IoT systems generate large amounts of data, which they must be able to manage while maintaining their integrity and accessibility. Due to the specific nature of IoT systems data, these systems typically use NoSQL databases such as Apache Casandra [18] or time-series databases such as InfluxDB [19] or TimescaleDB [20]. The system should also allow us to analyze the data using various techniques so that we can obtain statistics and trends or seasonalities. Finally, it is important to be able to present all these data in meaningful and easy-to-understand formats.
   At this point, although Home Assistant may be a very good solution for simpler scenarios, it had some limitations for our intended use cases. First, Home Assistant stores the collected sensor data in its own SQL database, but it is not intended to store large amounts of data in the long term. For this reason, we decided to use an external dedicated database. In our case, we deployed an InfluxDB 2.0 database [19]. To store the sensors data in the InfluxDB database, Home Assistant already includes an InfluxDB integration, which makes it easy to transfer the desired data from Home Assistant to the InfluxDB database. Another limitation comes with respect to the data analysis and visualization. Home Assistant supports dashboards with different panels to visualize data, but they are only intended for simple cases and have several limitations. Moreover, since we will use an external database to store the long-term data, we need a solution to visualize the data from this database. Therefore, we decided to use an external dedicated data visualization software. In our case, we deployed an instance of Grafana [21], which we configured to query to the InfluxDB database.

4. **Availability**: Availability is one of the key challenges of any computer system. In the context of an IoT platform or application, it means that it should be available to devices and users. For critical or enterprise applications, a minimum availability, calculated as the ratio of system uptime to total time, is typically required. In these cases, systems must be designed to eliminate single points of failure, using redundant systems with fault detection capabilities. In our case, our system was not intended to be used by critical or external users (outside the university) who would require high availability. Moreover, since we deployed the system in virtual machines running on the modern data center at the CITIC, a very low probability of failure is expected.

5. **Scalability**: As the number of IoT devices grows, the amount of data generated can quickly become overwhelming. Therefore, it is important to ensure that the system resources are appropriately scaled to handle this data, and that more resources can be easily added when needed. There are two common scaling strategies: (*1*) vertical scaling, which simply adds more resources to the current machines (e.g., memory or CPU cores), and (*2*) horizontal scaling, which adds more machines. The vertical scaling strategy is the simplest, since it only requires upgrading the current machines. On the other hand, the horizontal scaling strategy is more complex to maintain and operate, since it requires that the deployed software is designed to run on multiple nodes concurrently, and also requires additional software to manage the different nodes.

    In the case of our system, both Home Assistant and InfluxDB (in its open source version) do not support horizontal scaling, so vertical scaling is our only option to handle additional data and processing needs. However, this approach is appropriate for small to medium-sized IoT systems like ours. In particular, in Section 5 we show statistics of the central processing unit (CPU) and data usage of our system, where it can be seen that our system has several times more processing power and storage capacity than the current needs of the system.

6. **Maintenance**: IoT devices, like any other technology, need to be maintained. This includes activities such as cleaning, updating firmware, replacing batteries, and troubleshooting any problems that may arise. In particular, whenever possible, we should monitor the proper functioning of the devices and report any situation requiring maintenance. More specifically, for wireless devices, battery status and connectivity are important parameters to monitor. In wireless networks, problems can easily occur, leaving the devices disconnected.

    For this purpose, in our system we used both Grafana and Node-RED [22]. Grafana, as explained earlier, allows us to provide different visualizations of the data in our system, so we were able to plot different real-time sensor data such as battery levels or received signal strength indicator (RSSI) levels. Node-RED is an automation tool that allows us to easily create flow-based applications. Using Node-RED, we created several flows to monitor battery levels of several devices and send an alarm when certain thresholds were exceeded. We show some more general examples of monitoring with Grafana and Node-RED in Section 6.

7. **Security and privacy**: IoT devices may collect data about people and their environment, have access to sensitive data, or can be used to control critical systems. Therefore, it is important to ensure that they are secure and that data is kept private. Security and privacy measures include protecting devices from physical tampering, as well as securing communication channels and data storage.

    A common measure to protect communication channels is to use encrypted protocols to prevent unauthorized parties from intercepting and reading the data. Therefore, in our case, we used encrypted protocols such as HTTPS for all services and devices directly exposed to the Internet. However, in some cases we had to deal with devices that only used unencrypted communications protocols such as BACnet (Section 4.3), or Modbus/TCP (Section 4.4). In these cases, the devices were connected to a virtual local area network (VLAN) within the university's local area network (LAN). A VLAN allows for the creation of a separate network within a larger network that can be used to segment traffic and restrict access to specific devices or users. This helps to limit the potential attack surface of a system and prevents unauthorized parties from accessing the network and devices.

    In addition to protecting communication channels, data storage was also protected by implementing appropriate access controls and regular backups. In particular, permissions for users should only allow read access to data, and should restrict the data that can be accessed, e.g., only allowing access to a subset of the data. These measures prevent unauthorized access to data, and ensure that data can be recovered in the event of a failure or a security breach.

## 3. Specific interoperability and integration challenges

In IoT systems, several challenges need to be considered when integrating different devices and systems. When acquiring new devices for an IoT system, they should use open standards and flexible architectures that can be easily integrated into the system. However, in many cases we may want to integrate some previously deployed devices, and in these cases compatibility issues may arise. In addition, some other problems may arise unexpectedly during the integration process. In the literature, there exist several works detailing different real-world IoT solutions and systems, [23–34], however, these works usually consider only a single type of device or do not address the specific challenges that may be encountered when integrating different devices into IoT systems. Instead, in this work we consider the integration with different types of devices and systems, and below we describe in a comprehensive way the specific interoperability and integration challenges that we found during the integration process of such systems. Later in Section 4 we will describe these systems in detail.

1. **Sensor data not directly available**: In some cases, instead of obtaining the data directly from an IoT sensor, such data are retrieved from a cloud system to which the sensor is connected. This may be undesirable for security, privacy, and reliability reasons. In addition, users are exposed to changes (technical or policy) in the web services provided by the cloud company. However, in some cases this may be the only viable option to collect the sensor data. In this work we will show examples of this situation in Sections 4.1 and 4.4.2.

2. **Software licenses**: In some cases, licenses are required to use certain services or features of IoT devices. These licenses may be temporary or may be perpetual. On the one hand, temporary licenses allow for usage of services or features only for some amount of time, after which they should be renewed or a new license should be acquired. On the other hand, perpetual licenses grant access to the specific service or feature indefinitely. Perpetual licenses were traditionally used for software applications to be run on local devices or machines, although in recent years temporary licenses are being used more and more for these kinds of applications. Temporary licenses are also typically used for cloud software as a service (SaaS) applications. In this work, we needed to consider the usage of licenses for the scenario in Section 4.1.

3. **Customer Support**: Sometimes we may have questions about how a particular product works, its features, or how to use it, or we may experience some technical issues and need help resolving them. In these cases, we may need to contact the company's customer support.

   Customer support refers to the services provided by a company to its customers to assist them with product or service-related inquiries, issues, and problems. This can include various channels such as phone, email, chat, social media, and more.

   In this regard we may distinguish between free and paid customer support. Free customer support is often more basic and may not include personalized support, priority service, or quick resolution of issues. Paid support, on the other hand, is a type of support service that customers pay for. This type of support typically provides customers with more comprehensive and personalized support, including priority service and rapid resolution of issues. Paid customer support is often offered to customers as an additional service or as part of a premium package. Therefore, when purchasing some sensors or solutions from a company for an IoT platform, it is important to evaluate the types and prices of the customer support services. In this work, customer support services were needed for the integrations shown in Sections 4.2 and 4.4.2.

4. **Monitoring of wireless devices**: As mentioned in the Maintenance Item in Section 2, it is important for the case of wireless devices to monitor their battery status and connectivity. We show in Sections 4.1 and 4.2 two scenarios where wireless devices were used and we needed to monitor their battery levels and connectivity status. This turned out to be important for the scenario in Section 4.2, where we found some problems in several devices.

5. **Usage of gateways**: Gateways are required in IoT systems for devices or sensors that use network technologies that cannot be directly connected to the internet or the IoT platform. In such cases, a gateway acts as a bridge between the devices and the IoT platform, allowing them to communicate with each other. The gateway may also perform additional functions such as data processing, protocol translation, and data aggregation before forwarding the data to the IoT platform.

   Gateway devices are used in many of the integrations described in Section 4. In some cases they may be provided by the same companies that provide the sensors, as shown in Sections 4.1 and 4.2, or in other cases they may simply be commercial off-the-shelf (COTS) devices, such as the BACnet controller shown in Section 4.3. However, a challenge arises in some other cases where no suitable COTS gateway is commercially available. In such cases, it may be necessary to develop custom gateways, as will be shown in Sections 4.2 and 4.3.

   In addition, when dealing with gateways for devices using wireless technologies, we must consider the coverage and connectivity challenges associated with these technologies. Therefore, it is important to carefully plan the location of the gateways to ensure that they can effectively reach and communicate with all the sensors. Depending on the deployment scenario, multiple gateways may be required to increase the coverage. This was the case for the scenarios in Sections 4.1 and 4.2. Furthermore, once that the gateways and devices are deployed, gateways should provide self monitoring data to check for any possible connectivity issues that may occur with the devices, such as weak signal or signal interference. In particular, in Section 4.2 we show how it was decided to replace the originally deployed gateways with other custom-developed ones in order to debug some network issues.

6. **Protocol implementation not readily available**: When working with an IoT platform, different devices may use different communication protocols. However, some protocol implementations may not be readily available in the platform, which can make it difficult or impossible to integrate those devices with the platform. To overcome this problem, developers may need to do additional work to add these protocol implementations to the platform or to deploy additional gateways or proxies that allow communication between these devices and the IoT platform. In our case, we found this problem when integrating the devices shown in Section 4.3, which use the BACnet protocol. Currently, Home Assistant does not have a BACnet integration available, so we had to add it ourselves. This is discussed in Section 4.3.

7. **Proprietary and undocumented protocols**: Some devices may use proprietary communication protocols developed by the device manufacturer. In many cases they may not provide documentation or specifications for the protocol, which can make it difficult or impossible for other devices or systems to communicate with the proprietary device. This means that it may be not possible to integrate the device into an IoT system, leading to vendor lock-in, where the customer is dependent on the original manufacturer for support, upgrades, or replacements. In this work, we show an example of a device using a proprietary undocumented protocol in Section 4.3, in this case the protocol was simple and we were able to reverse-engineer the protocol and communicate with the device, but in other cases this may not be feasible.

8. **Protocol implementation with limitations or missing certain features**: In an IoT platform, a protocol implementation may have some limitations or lack certain features that are required for the specific use case or application. This can make it difficult or impossible to use the protocol for the intended purpose. In this work, we show an example of this in Section 4.4.2 for the Modbus/TCP protocol. In the Home Assistant platform, when using the Modbus/TCP protocol a TCP connection between the platform and the device is created to read the required data, but for efficiency reasons this connection remains open and cannot be closed. Devices implementing the Modbus/TCP protocol typically only allow a single connection at a time over TCP, this means that if a connection has already been established to the device, no other device or system can connect to it until the current connection is closed. This is a limitation in some cases where we want to access the same Modbus/TCP device from different client devices at different times. In the specific case shown in Section 4.4.2, this limitation could not be assumed and we had to add an alternative Modbus/TCP implementation to the platform.

**Table 1**
Summary of the list of challenges marking for each one the specific sections devoted to the scenarios where the challenge occurred.

| Challenge | Sections | | | | |
|---|---|---|---|---|---|
| | 4.1 | 4.2 | 4.3 | 4.4.1 | 4.4.2 |
| 1. Sensor data not directly available | ✓ | | | ✓ | |
| 2. Software licenses | ✓ | | | | |
| 3. Customer Support | | | ✓ | | ✓ |
| 4. Monitoring of wireless devices | ✓ | ✓ | | | |
| 5. Usage of gateways | ✓ | ✓ | ✓ | | |
| 6. Protocol implementation not readily available | | | | ✓ | |
| 7. Proprietary and undocumented protocols | | | | ✓ | |
| 8. Protocol implementation with limitations or missing certain features | | | | ✓ | ✓ |
| 9. Non-standard or faulty protocol implementation | | | | | ✓ |

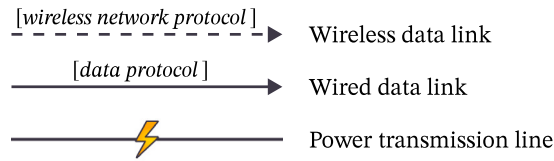

**Fig. 2.** Notation used for the diagrams shown in Section 4.

9. **Non-standard or faulty protocol implementation**: It is possible that some devices may implement protocols that do not conform to the standard, or that the implementation may contain some software bugs. This can lead to compatibility issues, where the device may not be able to interact with other devices or systems, and errors or unexpected behavior in the device or system, which can lead to unreliable or incorrect data, or malfunction of the device or system. In this work, we show an example of this in Section 4.4.2, where we tried connecting to a device using the Modbus/TCP protocol and experienced several unexpected connection errors, which we believe are due to a non-conforming or faulty protocol implementation.

Finally, in Table 1 we summarize the list of challenges, indicating for each one the specific sections devoted to the scenarios in which the challenge occurred.

## 4. Sensors and systems integrated into the IoT platform

In the following sections we will describe and analyze the integration of the platform with the different scenarios previously mentioned in the introduction. Several diagrams will be used to illustrate the platform integrations. These will use the notation shown in Fig. 2. Thus, a dashed arrow will indicate a wireless data link (with the wireless network protocol indicated in the text above), a solid arrow will indicate a wired data link (with the data protocol indicated in the text above), and a solid line with a lightning symbol will indicate a power transmission line. Finally, the direction of the data link arrows indicates the direction of the data transmission in the integrations.

### 4.1. University-wide real-time monitoring of $CO_2$

As mentioned in the introduction, we deployed the presented IoT system in early 2020 with the intention of limiting the spread of the SARS-Cov-2 virus in our university by ensuring good air quality. For this, we implemented real-time monitoring of $CO_2$ levels in different indoor spaces of our university, similar to what was done in other smaller educational institutions (preschool, primary, and secondary schools) [35]. To measure the $CO_2$ concentration in real time, it was necessary to deploy a network of sensors and monitor its values in the different spaces to be controlled. In the case of our university, the first objective was to sample representative areas (classrooms, study areas, offices, etc.) to check the suitability of the ventilation conditions recommended by the authorities (i.e., the World Health Organization (WHO) and the regional government).

Due to the large spacing between the areas to be monitored, the "Aranet4 PRO" sensor [36] was chosen to keep the deployment simple. The Aranet4 PRO is a compact battery-powered device manufactured by Aranet. This device can not only transmit the sensor data to a nearby computer via Bluetooth, but also to an "Aranet PRO base station" [37] up to 3 km away via the LoRa [38] technology using the EU868 band (868 MHz for Europe) or the NA915 band (915 MHz for North America). In both cases, the Aranet4

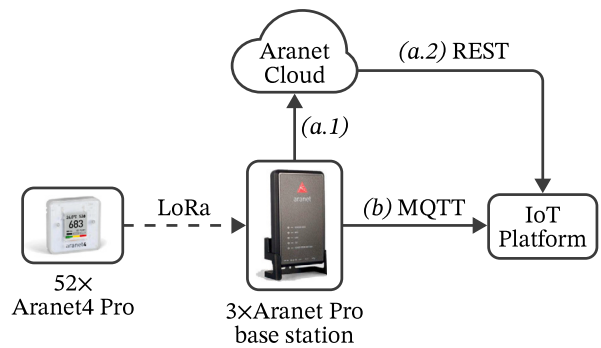

**Fig. 3.** Integration diagram of the Aranet4 Pro devices.

PRO device can be configured to transmit the sensor data in 1 min, 2 min, 4 min, or 10 min intervals. The Aranet4 PRO is one of the $CO_2$ sensors that is recommended by the authors in [39], it provides a non-dispersive infrared radiation absorption (NDIR) sensor for $CO_2$, and also sensors for temperature, relative humidity, and atmospheric pressure. For this work, we used 52 Aranet4 PRO devices and 3 Aranet PRO base stations.

The Aranet PRO base stations can be connected to a LAN via Ethernet or Wi-Fi. The sensor data collected in the base stations can then be retrieved in the following ways:

- *Message Queuing Telemetry Transport (MQTT)*: The Aranet PRO base stations can be configured to publish data to a broker using the MQTT protocol. In this way, data can be obtained by subscribing to the corresponding topics of the data sent to the broker.
- *Aranet Cloud*: The manufacturer of the devices (Aranet) also provides a cloud service called "Aranet Cloud", to which the Aranet PRO base stations can upload the sensor data. In this way, the sensor data can be retrieved from the Aranet Cloud using a REST API.

In Fig. 3 we show the diagram of the integration of the Aranet4 Pro devices with the IoT platform. The Aranet's business model for its cloud is based on temporary licenses. In a first phase, licenses for this cloud were considered. The reason for this decision was that, at the time these devices were acquired, we had not yet deployed our platform. So, once the platform was deployed, we configured it to periodically collect the data via the Aranet cloud REST API. This is shown in Fig. 3 with the links labeled *(a.1)* and *(a.2)*. As soon as it was possible, the cloud data collection was replaced by MQTT. This required the purchase of three licenses, since in this case the Aranet's business model consists in purchasing a perpetual license per base station. Then the data could be easily collected in Home Assistant using its MQTT integration. This is shown in Fig. 1 with the link *(b)*. This way we obtain the following six parameters from each Aranet4 device: $CO_2$ (ppm), temperature (°Celsius), humidity (%), atmospheric pressure (hPa), battery (%), and RSSI (dB).

Therefore, during the deployment and integration of the Aranet4 devices with our IoT platform we found the following specific interoperability and integration challenges, as described previously in Section 3:

- Sensor data not directly available: In a first step data were obtained from Aranet Cloud.
- Software licenses: We purchased Aranet Cloud and Aranet4 Pro MQTT licenses.
- Monitoring of wireless devices: The Aranet4 Pro devices are wireless devices running on batteries.
- Usage of gateways: Aranet Pro base stations.

Thanks to the availability of the platform and this integration, it was possible in 2021 to monitor and collect air quality data for a major event: the university entrance exams. In Spain, students who want to go to university must pass the university entrance exams, which are organized by the governments of the different Autonomous Communities of Spain.[2] These are written exams, with a duration of 90 min each, held over a period of three-days in the different universities of the Autonomous Communities, and to a lesser extent, in high schools. For the case of the Autonomous Community of Galicia (where the UDC is located), the university entrance exams in 2021 were celebrated between June 8 and June 10. In the UDC, the exams were distributed in 7 buildings in three different campuses, located in two cities 70 km apart (A Coruña and Ferrol). A base station was installed in each of the campuses, which was sufficient to cover the classrooms of interest thanks to the coverage range supported by LoRa. In addition, we have published the dataset of these measurements [6].

As a summary, Fig. 4 shows the time evolution of the $CO_2$ statistics during the three days considering all the monitored rooms. An increase of the $CO_2$ concentration can be observed during the celebration of the exams. The advantage of collecting these data in such a controlled situation is that they can be analyzed to extract mathematical models that relate rooms, occupancy, and ventilation, so that they can be extrapolated to other rooms without the need for additional measurements [7,8].

---

[2] In Spain, Autonomous Communities are the first-level political and administrative divisions, with self-government and legislative powers, recognized by the Spanish Constitution of 1978.

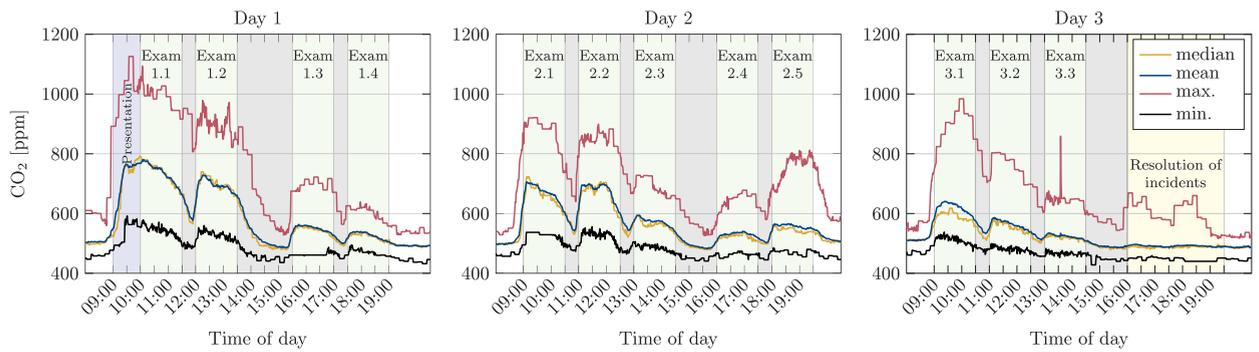

**Fig. 4.** Time evolution of $CO_2$ statistics for the three days of the University entrance exams. In each graph, the time slots for each exam are identified. The last day there is a time slot for solving any issues related to the exams, which is labeled as "Resolution of incidents".

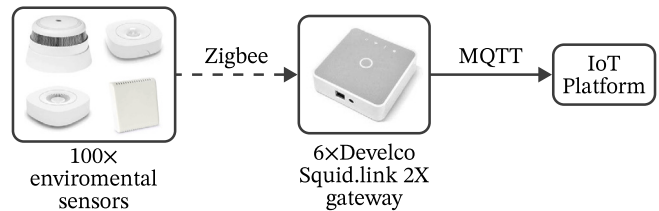

**Fig. 5.** Integration diagram for the CITIC building monitoring.

### 4.2. Building-wide real time monitoring of various parameters

Shortly after deploying the $CO_2$ monitoring system described in the previous section, the management of the CITIC research center decided to deploy a monitoring system in their building. As stated in the introduction, the CITIC is a research center of the University of A Coruña (UDC). The management of the center wanted a system to monitor several environmental parameters such as temperature, humidity, and air quality, as well as the occupancy of the different spaces in the building. In this case, the deployment of the sensors and gateways was carried out by the UDC spin-off Sevenix (Sevenix Ingeniería S.L.)[3], and we were tasked with the integration of the sensors into our platform to monitor the data generated.

Since the area to be monitored was limited to a single building, sensors using the Zigbee technology were chosen. Zigbee is a low-power, low data rate communication technology that uses a multi-hop mesh topology [9,40], with a device range of 10 m to 100 m. Although this is a smaller range that the one provided by the LoRa technology discussed in the previous section, the range of a Zigbee network can be easily extended by adding new routers. In addition, there are many sensors and IoT devices on the market that use Zigbee as their communication technology, making this technology a good solution for small or medium-sized deployments. In particular, the devices that we used in this case were the following:

- Develco "Squid.link 2X" gateway (DP-44): Zigbee coordinator to which the sensors send their messages. When a message arrives, it is sent back to our platform via MQTT.
- Develco air quality sensor (AQSZB-110): Provides a measure of the air quality by measuring the volatile organic compounds (VOCs) levels. It also provides ambient humidity and temperature parameters.
- Develco humidity sensor (HMSZB-110): Provides ambient humidity and temperature parameters.
- Develco motion sensor (MOSZB-140): Sends an occupancy message when it detects motion. It also provides ambient light and temperature parameters.
- Develco smoke alarm (SMSZB-120): Sends a smoke alarm message when it detects smoke. It also provides an ambient temperature parameter.
- Titan Products $CO_2$, humidity, and temperature sensor (TPZRCO2HT-Z3): Measures ambient $CO_2$ levels, humidity, and temperature.

The CITIC research center building consists of 6 floors. In each one a Zigbee network was deployed using the Develco "Squid.link 2X" gateway as the coordinator. For most of the floors, the total area to be monitored was too large and the coverage of this gateway was not sufficient. In such cases, additional Zigbee routers were placed to adapt the coverage of the communication networks. The routers used were Develco Smart Siren (SIRZB-110) devices. These devices combine the functionalities of a siren, a voice prompt, and a Zigbee router, although in our case we only use them as routers.

---

[3] http://www.sevenix.es/

**Table 2**
CITIC building deployment data per floor of the area monitored, the number of Zigbee routers, and the number of sensors.

| Floor | Monitored area (m²) | Zigbee routers | Number of sensors | | | | | |
|---|---|---|---|---|---|---|---|---|
| | | | Air quality | Motion | Humidity | Smoke | $CO_2$ | Total |
| −2 | 71.3 | 0 | 1 | 2 | 0 | 1 | 1 | 5 |
| −1 | 67.6 | 1 | 0 | 3 | 2 | 0 | 0 | 5 |
| 0 | 526.9 | 2 | 5 | 21 | 1 | 3 | 3 | 33 |
| 1 | 420.7 | 2 | 6 | 16 | 0 | 2 | 5 | 29 |
| 2 | 417.3 | 1 | 2 | 18 | 0 | 2 | 2 | 24 |
| 3 | 26.0 | 0 | 0 | 2 | 1 | 1 | 0 | 4 |

**Table 3**
Parameters received from the CITIC devices. Parameters that are transmitted by all the devices are listed in the "Common parameters" column, other specific parameters are listed in the "Specific parameters" column. The unit (for numeric parameters) or data type (for non-numeric parameters) is specified within parentheses.

| Device | Specific parameters | Common parameters |
|---|---|---|
| Develco air quality sensor | • Battery low (boolean)<br>• Humidity (%)<br>• VOC (ppb)<br>• VOC level (text)<br>• VOC recommendation (text) | |
| Develco humidity sensor | • Humidity (%) | Battery voltage (V)<br>Network link strength (%)<br>Temperature (°C) |
| Develco motion sensor | • Alarm (boolean)<br>• Alarm tamper (boolean)<br>• Alarm trouble (boolean)<br>• Battery low (boolean)<br>• Battery defect (boolean)<br>• Illuminance (lx)<br>• Occupancy (boolean) | |
| Develco smoke alarm | • Battery low (boolean)<br>• Smoke alarm (boolean) | |
| Titan Products $CO_2$, humidity, and temperature sensor | • $CO_2$ (ppm)<br>• $CO_2$ level (text)<br>• Humidity (%) | |

In Fig. 5 we show the diagram of the integration with our platform. As shown in the diagram, each sensor sends its readings to its corresponding gateway, and then the gateway sends these data back to our platform using the MQTT protocol. In Table 2 we show, for each floor, the total monitored area (i.e., the combined area of the monitored rooms), the number of Zigbee routers, the number sensors by type, and the total number of sensors. In Table 3 we detail the parameters received from each device.

One problem found during this project was that the CO2 sensors were consuming more power than they should. This demonstrated one of the major weaknesses of this deployment: the closed nature of the Develco gateways meant that the Sevenix company could not debug the problem without support from the Develco company. The contract Sevenix signed with Develco included a limited amount of premium customer support, and after that any additional customer support should be paid for, but this is expensive for a small company such as Sevenix. Thus, in order to avoid this problem in the future, Sevenix decided to replace the Develco gateways with its own custom-developed Zigbee gateways. These gateways have been developed with an RF-STAR RF-BM-2652P2 multiprotocol module [41], they would allow easy debugging of any problems that may occur in the network, and they incorporate the possibility of a native integration with Home Assistant instead of using MQTT. In Fig. 6 we show these new custom-developed gateways.

Finally, we summarize the specific interoperability and integration challenges we encountered during the integration of the CITIC research center devices with our IoT platform, as previously discussed in Section 3. These challenges are:

- Customer Support: Customer support was needed for the Develco devices.
- Monitoring of wireless devices: All devices except the gateways were wireless devices running on batteries.
- Usage of gateways: First the Develco gateways were used, then the custom-developed Zigbee gateways.

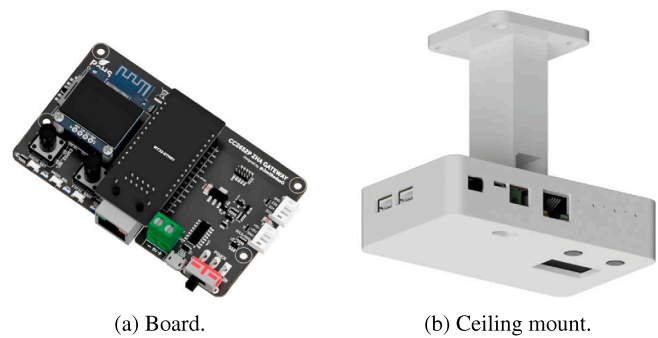

(a) Board.  (b) Ceiling mount.

**Fig. 6.** Custom-developed Zigbee gateways.

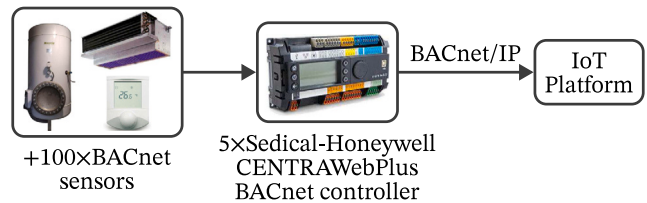

**Fig. 7.** Integration diagram for the HVAC, water heating, and radon sensors via the BACnet protocol.

### 4.3. HVAC, water heating, and radon sensors

Improving energy efficiency is one of the current key objectives of the European Union (EU) as part of the 2030 Agenda for Sustainable Development set by the United Nations General Assembly (UN-GA) [42], as energy savings allow saving money and reducing greenhouse gas emissions. Thus, recently, the architecture, urban planning and equipment department of the UDC carried out several modifications in different buildings to improve their energy efficiency. Among them, some of the most important were:

- Improvement of the HVAC systems.
- Installation of thermal insulation.
- Replacement of fluorescent lamps by LED ones.
- Installation of presence detectors.
- Installation of $CO_2$ environmental sensors.
- Installation of fan coil units (FCUs).

In addition, two radon sensors were installed in the Faculty of Computer Science of the UDC for health reasons. Radon (Rn-222) is a radioactive chemical element that belongs to the noble gas group. It is gaseous, colorless, odorless, denser than air, and soluble in water and other liquids. It comes from the decay chain of uranium (U-238) found in the rocks of the earth's surface. There are geographic areas in which, due to their geology, buildings may contain high concentration levels of radon, especially in basements or on the ground floor. Radon is one of the leading causes of lung cancer, and thus it is a public health problem [43]. According to the 2013/59/Euratom EU directive, as well as to the Spanish law (Royal Decree 783/2001), the radon level in workplaces must not exceed $300\,\text{Bq/m}^3$. Therefore, radon sensors allow for checking that the monitored buildings do not exceed these levels and, when necessary, apply forced ventilation systems, thus ensuring the safety of their occupants.

The BACnet protocol is used for collecting the data from the aforementioned sensors and monitoring systems. BACnet is a data communication protocol developed by the American Society of Heating, Refrigerating and Air-Conditioning Engineers (ASHRAE) [44] for building automation and control (BAC) networks, and published as both the American standard ANSI/ASHRAE 135 and the international standard ISO 16484-6. Today, it is the most widely used of the BACs network protocols, with a global market share of over 60%, according to a 2018 report by the Building Services Research and Information Association (BSRIA) [45].

In BACnet installations, the sensors and monitoring systems are connected to a BACnet controller. This is just a computer with which we can interact via the BACnet protocol to read the values of the desired sensors and systems. In our case, as shown in Fig. 7, the installed sensors and monitoring systems are connected to a corresponding Sedical-Honeywell CENTRAWebPlus BACnet controller. These BACnet controllers have an Ethernet port, which was used to connect them to the university's LAN. Then, we periodically poll the desired sensor data from our platform using the BACnet/IP protocol. In Table 4 we detail the BACnet deployments that were integrated into our platform, with the specific monitored parameters for each BACnet controller and the total number of parameters.

Unfortunately, Home Assistant does not have a BACnet integration readily available, and so we had to add it ourselves. For this, we used the existing "BACnet Stack" library [46], which provides an open-source BACnet protocol stack implementation. The "BACnet Stack" library already provides the demo applications `bacrp` (BACnet read property) and `bacrpm` (BACnet read property multiple) to send BACnet request messages to read one or multiple sensor values, respectively. However, these applications return the data in a non-standard format. So we modified the `bacrpm` application to return the data in a standard JSON format and we integrated it into Home Assistant. Next, we configured Home Assistant to periodically call the `bacrpm` application to poll the desired data from the sensors. This integration has been published to a public open source repository, as shown in Section 7.

**Table 4**
Details of the BACnet deployments. Rector's office has 2 BACnet controllers, whereas the other buildings have one each. For each BACnet controller, the specific monitored parameters and the total number of parameters are specified. If similar parameters are received for multiple rooms or zones, their number is indicated within square brackets. The unit (for numeric parameters) or data type (for non-numeric parameters) is specified within parentheses.

| Building | Monitored parameters | Total number of parameters |
| --- | --- | --- |
| Faculty of Computer Science | • Radon concentration [2 rooms] ($Bq/m^3$) | 2 |
| Sports Pavilion | • Ambient indoor temperatures [2 zones] (°C)<br>• Ambient outdoor temperature (°C) | 3 |
| "Normal" building | • Radon concentration [2 rooms] ($Bq/m^3$)<br>• Ambient indoor temperatures [8 rooms] (°C)<br>• Ambient outdoor temperature (°C)<br>• Energy used for air heating [2 zones] (MW h)<br>• Energy used for water heating (MW h)<br>• Total energy used for heating (MW h)<br>• Cubic meters of natural gas used for heating ($m^3$) | 16 |
| Rector's Office | BACnet controller 1:<br>  • Ambient indoor temperatures [38 rooms] (°C)<br>  • Ambient outdoor temperature (°C)<br>  • Fan coil units on/off state [38 rooms] (boolean) | 77 |
| | BACnet controller 2:<br>  • Ambient indoor temperatures [16 rooms] (°C)<br>  • Ambient $CO_2$ concentration [2 zones] (ppm)<br>  • Fan coil units on/off state [16 rooms] (boolean) | 34 |

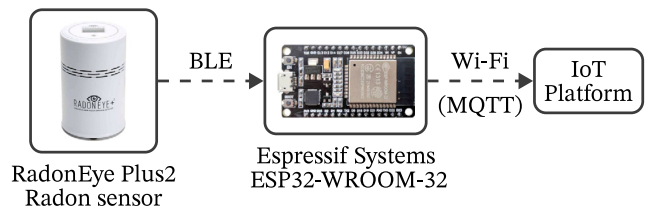

**Fig. 8.** Integration diagram for the RadonEye Plus2 radon sensor.

In addition to the radon sensors mentioned above, we have also integrated two low-cost RadonEye Plus2 (RD200P2) radon sensors from FTLAB Corp. [47] as shown in Fig. 8. As shown in Fig. 8, these sensors provide their readings via a Bluetooth Low Energy (BLE) connection. To integrate these sensors with our platform, we developed an external ad-hoc gateway device using an Espressif Systems ESP32-WROOM-32 module, which is based on an ESP32 chip [48]. This is a small and low-cost device that contains a programmable ARM processor along with numerous ports and input/output interfaces. We used this particular model because it has BLE support, which allows us to collect data from the radon sensor, and it also has Wi-Fi support, which allows us to send the results to our platform via MQTT. With this setup, we read three parameters from the RadonEye Plus2 sensors: radon level ($Bq/m^3$), temperature (°Celsius), and humidity (%). For the control software of the ESP32-WROOM-32 module, we used the ESPHome project [49]. This is an open source project that allows easy development of different types of sensors using ESP32-based modules. However, the ESPHome project did not have support for the specific RadonEye sensor model that we used (RD200P2), but only for the previous model (RD200). So we implemented this functionality and submitted a request to integrate it into ESPHome, as shown in Section 7.

Finally, we summarize the specific interoperability and integration challenges we encountered during the integration of the aforementioned devices with our IoT platform, as previously discussed in Section 3. These challenges are:

- Usage of gateways: BACnet gateways and our custom-developed ESP32-WROOM-32 gateways for the RadonEye Plus2 sensors.
- Protocol implementation not readily available: We had to add support for the BACnet protocol to our IoT platform.

### 4.4. Energy monitoring

As part of the university's energy efficiency efforts, the architecture, urban planning and equipment department needed access to energy data from multiple systems at the university. In particular:

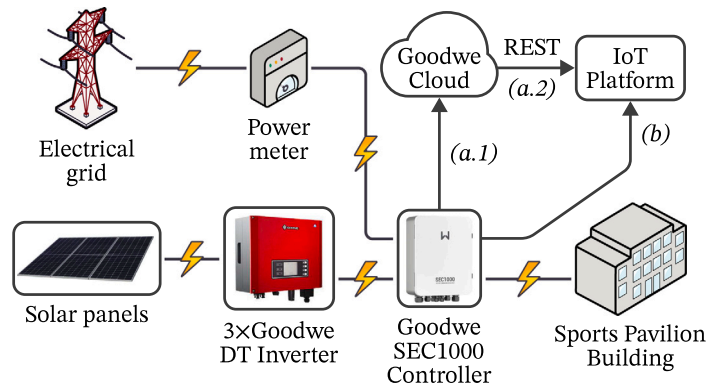

**Fig. 9.** Solar power system for the Sports Pavilion building.

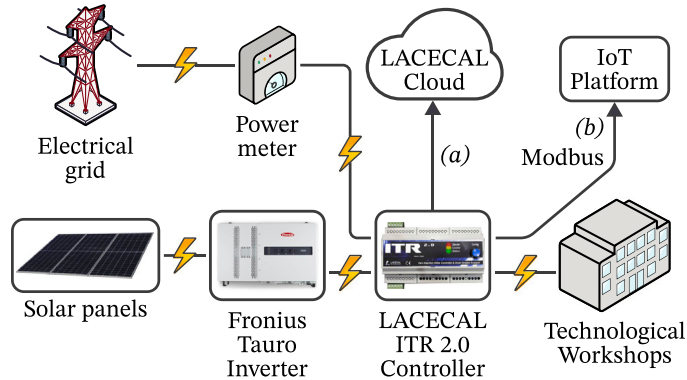

**Fig. 10.** Solar power system for the Technological Workshops building.

- Monitoring of solar energy systems in two buildings. The systems were integrated into our platform to monitor several of their parameters, allowing the department's staff to verify their correct operation and the energy savings achieved. This case will be detailed in Section 4.4.1.
- Monitoring the energy consumption of the electrical network of several buildings and facilities. To do this, we needed to retrieve data from various energy meters connected to the university's LAN via Ethernet interfaces. This case will be detailed in Section 4.4.2.

#### 4.4.1. Solar energy monitoring

Photovoltaic solar power systems were installed in two buildings of the UDC to further improve their energy efficiency, one in the Sports Pavilion building, and the other one in the Technological Workshops building. The architectures of these systems are shown in Figs. 9 and 10. They are similar, although the devices used are from different vendors in each case. Such systems consist of *(1)* photovoltaic solar panels, *(2)* a set of inverters, and *(3)* an energy controller. In both cases, the solar panels were first installed on the roofs of the buildings to convert the sunlight into direct current (DC). Second, inverters were installed inside the buildings to convert the DC generated by the solar panels into alternating current (AC) compatible with the electrical grid. Finally, an energy controller was installed in each building to monitor the power consumed in the building, and based on this, to control the power produced by the solar power system.

In the case of the solar power system of the Sport Pavilion building shown in Fig. 9, the installed inverters were Goodwe GW15KN-DT, each one capable of producing up to 15 kW of power, and the energy controller was a Goodwe SEC1000 [50]. Such an energy controller was connected to the university's LAN and the Internet via an Ethernet interface. Then, as shown in Fig. 9, we were able to retrieve the monitored data in two ways:

1. The Goodwe SEC1000 was connected to the manufacturer's cloud platform, to which it transmitted the system's energy and power data. This data could be then retrieved from the cloud via a REST API. In Home Assistant, there already exists an integration for this device developed by the community of Home Assistant developers [51], which allows to easily retrieve the desired data. Hence, as a first step, we used this method to integrate the system into our platform as shown in Fig. 9 with the labels *(a.1)* and *(a.2)*.

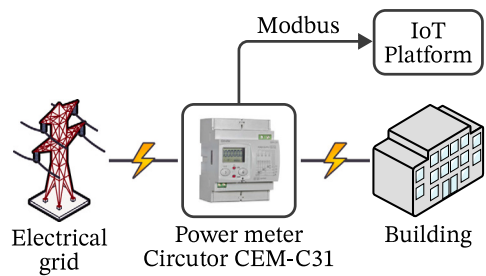

**Fig. 11.** Integration of the Circutor CEM-C31 power meters.

2. The manufacturer offers a software called "ProMate" which can be used to retrieve data from the device. Unfortunately, the manufacturer does not provide documentation about the protocol used for this communication. However, by capturing and examining the packets exchanged between the SEC1000 and the "ProMate" software, we were able to implement a Python script that was able to request and collect the same data shown in the "ProMate" software. Then, in Home Assistant, we periodically polled the SEC1000 energy controller with the implemented script. This integration is shown in Fig. 9 with the label *(b)*, and has been published to a public open source repository, as shown in Section 7.

Note, however, that the data obtained with these two methods are not the same. From the cloud, we can retrieve data of the energy and power produced by each individual inverter, but from the SEC1000 device we can only retrieve the accumulated energy and power for all inverters. Nevertheless, we prefer this second method because it does not require us to depend on an external service.

In the case of the solar power system of the Technological Workshops shown in Fig. 10, a Fronius Tauro inverter was installed, and the energy controller was a LACECAL ITR 2.0 device [52]. As in the previous case, this energy controller was also connected to the university's LAN and to the Internet via an Ethernet interface, and the device was connected to the manufacturer's cloud, to which it transmitted the system energy and power data. This is shown in Fig. 10 with the label *(a)*. Unlike the previous case, the manufacturer does not provide an API for its cloud system, so the data cannot be easily retrieved from it. However, in this case the data can be recovered directly from the energy controller via the Modbus/TCP protocol. Modbus/TCP [53] is a communication protocol used for industrial automation and control systems. It is a variant of the Modbus protocol that uses TCP/IP as the underlying communication layer instead of the serial communication protocols used in the original Modbus. Modbus/TCP is widely used in industrial automation, building automation, and other applications where reliable, low-cost communication is required. It is an open standard and is supported by many manufacturers and vendors, making it relatively easy to integrate with existing systems.

We have therefore integrated this device into our platform via the Modbus/TCP protocol, as shown in Fig. 10 with the label *(b)*. However, this integration was not straightforward. The usual process for retrieving data from a device via the Modbus/TCP protocol is with simple reading requests, but in this particular case, the LACECAL device provided some of the desired data as historical data, which required the following additional steps to retrieve it:

1. Write the date of the desired data (3 bytes starting at address $0 \times 1000$).
2. Wait for the requested data to be ready by checking the value of the register at address $0 \times 1003$.
3. Read the requested data from the registers.

Thus, we developed a custom code that performed these steps to integrate the device into our platform. This integration has been published to a public open source repository, as shown in Section 7.

Finally, we summarize the specific interoperability and integration challenges we encountered during the integration of the aforementioned devices with our IoT platform, as previously discussed in Section 3. These challenges are:

- Sensor data not directly available: Data from the individual inverters in the Sports Pavilion building were only available from the Goodwe Cloud.
- Proprietary and undocumented protocols: The protocol of the Goodwe SEC1000 Controller is not documented.
- Protocol implementation with limitations or missing certain features: We had to add an alternative Modbus/TCP implementation to the IoT platform to retrieve data from the LACECAL device.

*4.4.2. Grid energy monitoring*

Several Circutor CEM-31 power meters with Ethernet interfaces were installed in different buildings of the university and were connected to the university's LAN. These devices provided their instantaneous power readings via the Modbus/TCP protocol. This allowed us to easily integrate these devices into our platform by periodically reading the appropriate Modbus registers from the devices, in order to do this we used the Python PyModbus library [54], which we had also used for the previous integration of the LACECAL energy controller described in Section 4.4.1. This integration is shown in Fig. 11.

Another interesting energy meter was the one that monitors the energy consumption of the UDC Esteiro campus, located in the city of Ferrol. This power meter was a Circutor CIRWATT B device provided and installed by the electricity supplier company, and also connected to the university's LAN via an integrated Ethernet interface. This integration is shown in Fig. 12. According to the device manual, the instantaneous power data readings could be obtained from the device using the Red Electrica de España (REE, Electrical Grid of Spain) protocol. REE is a public limited corporation that operates the national electrical grid in Spain. The REE protocol is defined by REE based on the international standard IEC 870-5-102, and it is used by the Spanish electricity supplier companies for remote meter reading.

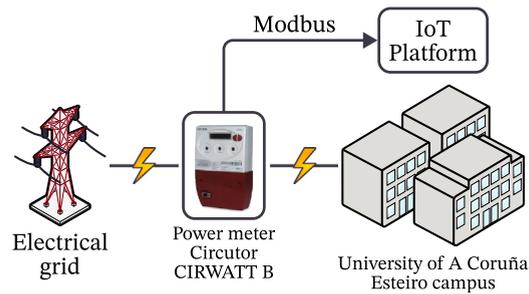

**Fig. 12.** Integration of the Circutor CIRWATT B power meter.

Since we were not familiar with the REE protocol, and in order to get some more details about the power meter, we contacted Circutor's technical service, which informed us that the data could also be retrieved via the Modbus/TCP protocol (this was not mentioned in the device manual). Furthermore, the Modbus/TCP service was enabled on the port number 10000 instead of 512, which is the port number assigned by the Internet Assigned Numbers Authority (IANA) for the Modbus/TCP protocol. Unlike the REE protocol, the Modbus/TCP protocol is a very simple protocol and we were familiar with it, so we decided to use it to retrieve the desired data from the device. However, we encountered some unexpected problems:

1. First, we tried querying the device with `mbpoll` [55]. `mbpoll` is a command line utility that allows to easily read and write data to and from devices via Modbus. We had previously used successfully this tool for initial testing of several other devices, but with the Circutor CIRWATT B device the `mbpoll` tool always failed with a connection error.
2. Second, we tried to query the device using the Python PyModbus library [54], which we had used for the Circutor CEM-C31 power meters described above and the LACECAL energy controller described in Section 4.4.1. The PyModbus library, in versions 3.0.x, has both a synchronous and an asynchronous implementation. These two implementations should in principle be the same from the point of the server side, being different only for the client, but in practice this turned out not to be the case. We first tried the PyModbus asynchronous implementation and only about half of the times a request was made data were returned successfully, while the other half resulted in a connection error. Then, we tried the synchronous implementation, in this case no errors occurred during all the tests we conducted, and all the data were returned successfully.

We were not able to determine the cause of the connection errors, but we suspect that it was due to some problem in the Modbus/TCP protocol implementation in the power meter device. Since the PyModbus synchronous implementation worked correctly we used it to integrate the device with our platform by periodically retrieving the device's instantaneous power readings.

It should be also noted that the usage of our own PyModbus Modbus/TCP implementation instead of the Home Assistant Modbus/TCP implementation was also motivated by the fact that we wanted to be able to access the Circutor devices from different client devices at different times. The aforementioned Circutor devices, as is typical in other Modbus/TCP devices, only allow a single connection at a time. However, the Home Assistant Modbus/TCP implementation does not close TCP connections for efficiency reasons, and thus it would not allow to access these devices from other clients. Conversely, our PyModbus Modbus/TCP implementation ensures that after each request the TCP connection is closed and thus other clients can access the devices.

Finally, we summarize the specific interoperability and integration challenges we encountered during the integration of the aforementioned devices with our IoT platform, as previously discussed in Section 3. These challenges are:

- Customer Support: Customer support was needed for the Circutor CIRWATT B device.
- Protocol implementation with limitations or missing certain features: With the Home Assistant Modbus/TCP implementation other clients could not access the devices.
- Non-standard or faulty protocol implementation: We found several connection errors with the Modbus/TCP protocol of the Circutor CIRWATT B device.

*4.4.3. Energy parameters monitored*

In Table 5 we detail the monitored parameters, as well as the total number of parameters for each of the energy systems described previously.

**Table 5**
Details of the monitored energy parameters. For each integration, the specific monitored parameters and the total number of parameters are specified. Note that three-phase electric power is used in these systems. Thus, for some measurements, such as the current or voltage, three parameters are monitored, one for each of the phases, this is indicated within square brackets. Additionally, for the power parameters, the sum of the powers of the individual phases may be also monitored. The unit, if any, is specified within parentheses.

| Integration | Monitored parameters | Total number of parameters |
| --- | --- | --- |
| Solar power system for the Sports Pavilion building (Fig. 9) | From Goodwe cloud (Fig. 9, *a.2*):<br>• Power produced by inverter [3 inverters] (W) | 3 |
| | From Goodwe SEC1000 controller (Fig. 9, *b*):<br>• Total power produced by the inverters (W)<br>• Power consumed by the building (W) | 2 |
| Solar power system for the Workshops building, LACECAL ITR 2.0 energy controller (Fig. 10) | Power:<br>• Consumed from the electrical grid [3 phases and total] (W)<br>• Consumed by the building [3 phases and total] (W)<br>• Produced by the inverter [3 phases and total] (W)<br>• Excess [3 phases and total] (W)<br>Energy (in 15-min periods):<br>• Total consumption [3 phases and total] (kW h)<br>• Consumption from the electrical grid [3 phases and total] (kW h)<br>• Exported to the electrical grid [3 phases and total] (kW h)<br>• Produced by the inverter (kW h) | 29 |
| Circutor CEM-C31 power meter (Fig. 11) | • Current [3 phases] (A)<br>• Voltage [3 phases] (V)<br>• Cosine of the phase [3 phases]<br>• Apparent power [3 phases and total] (V A)<br>• Active power [3 phases and total] (W)<br>• Reactive power [3 phases and total] (var)<br>• Imported energy (W h)<br>• Exported energy (W h) | 23 |
| Circutor CIRWATT B power meter (Fig. 12) | Same as the Circutor CEM-C31 power meter, and additionally:<br>• Frequency (Hz)<br>• Power factor<br>• Reactive energy per quadrant [4 quadrants] (var h) | 29 |

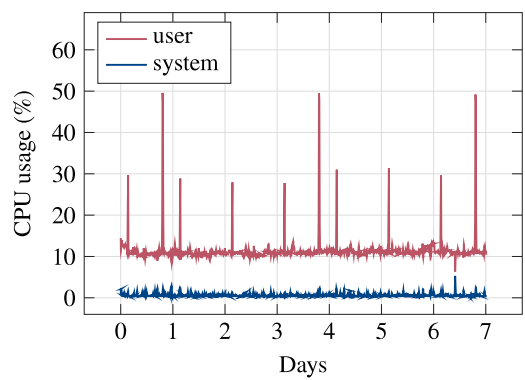

**Fig. 13.** CPU usage during a week collected from our IoT platform, averaged over 10-min intervals.

## 5. Data and CPU usage statistics

As mentioned in the Scalability Item in Section 2, one of the challenges in IoT systems is the large volume of data generated by the increasing number of devices. Therefore, it is important to ensure that resources are scaled appropriately. In this section we will analyze data and CPU usage statistics from our IoT systems with the integrations described in the previous section.

The IoT platform, as described in the Availability Item of Section 2, was deployed as a virtual machine in the CITIC data center. This virtual machine was assigned with two CPU cores at 2.20 GHz. In Fig. 13 we show the CPU usage of our IoT platform during a week averaged over 10-min intervals. It can be seen that there are a few peaks of CPU usage of 30% and 50%, but overall it stays fairly stable between 9% and 14%, indicating that the system is able to handle the data processing needs.

**Table 6**
Statistics of the data generated by the different devices and integrations described previously in Section 4.

| Section | Integration/Device | No. of devices | Param. per device | Total Param. | Avg. data points per hour | Avg. data points per hour and device |
|---|---|---|---|---|---|---|
| 4.1 | Aranet4 Pro | 52 | 6 | 312 | 2218.86 | 42.67 |
| 4.2 | Air quality sensor | 14 | 8 | 112 | 249.47 | 17.82 |
| | Motion sensor | 62 | 10 | 620 | 641.16 | 10.34 |
| | Humidity sensor | 4 | 4 | 16 | 19.53 | 4.88 |
| | Smoke sensor | 9 | 2 | 18 | 8.73 | 0.97 |
| | $CO_2$ sensor | 11 | 6 | 66 | 124.55 | 11.32 |
| 4.3 | BACnet controller (Faculty of CS) | 1 | 2 | 2 | 12.41 | 12.41 |
| | BACnet controller (Sports Pavilion) | 1 | 16 | 16 | 22.17 | 22.17 |
| | BACnet controller ("Normal" building) | 1 | 3 | 3 | 39.48 | 39.48 |
| | BACnet controller (Rector's Office 1) | 1 | 77 | 77 | 151.59 | 151.59 |
| | BACnet controller (Rector's Office 2) | 1 | 34 | 34 | 57.21 | 57.21 |
| | RadonEye Plus2 | 2 | 3 | 6 | 12.46 | 6.23 |
| 4.4.1 | Goodwe SEC1000 | 1 | 2 | 2 | 114.92 | 114.92 |
| | Goodwe cloud | 1 | 3 | 3 | 14.05 | 14.05 |
| | LACECAL ITR 2.0 | 1 | 29 | 29 | 49.03 | 49.03 |
| 4.4.2 | Circutor CEM-C31 | 12 | 23 | 276 | 19 065.15 | 1588.76 |
| | Circutor CIRWATT B | 1 | 29 | 29 | 2206.53 | 2206.53 |
| | **Total** | 173 | 257 | 1621 | 25 007.30 | – |

Regarding the data needs, the InfluxDB database was deployed on another virtual machine assigned with a disk capacity of 100 GB. In Table 6 we show different statistics of the data generated by the different devices and integrations that were described earlier in Section 4. The columns in this table are the following:

1. *Section*: The specific section in the paper where the integration or device specified in the next column (*Integration/Device*) is described.
2. *Integration/Device*: The name of the specific device model (e.g., Aranet4 PRO), or the integration (e.g., Goodwe cloud). Note that in the case of the BACnet controllers of Section 4.3, although they are the same device model, different data are obtained for each one. So we list each one separately with its location.
3. *No. of Devices*: The number of devices specified in the previous column that have been integrated into our system.
4. *Param. per Device*: The number of parameters transmitted by a device. The Aranet4 PRO devices, as stated in Section 4.1, transmit 6 parameters: temperature, humidity, pressure, $CO_2$ level, RSSI level, and battery parameters. For the other devices these numbers were also defined in Tables 3–5.
5. *Total Param.*: The total number of parameters considering all the devices, this is just the result of multiplying the corresponding *Param. per Device* data by the *No. of Devices* data.
6. *Avg. Data Points per Hour*: A data point is a single piece of information or measurement record that is collected and recorded; more specifically, in our case, it is a measurement record from a device (e.g. temperature, humidity, etc.) that is associated with a timestamp and may also include additional metadata (e.g. device that took the measurement, its location, etc.). This column shows how many data points per hour are recorded into the InfluxDB database (see Fig. 1) considering all the devices (i.e., the number of devices stated in the column *No. of Devices*). The results shown in the table were obtained by averaging over all the data points collected during a one-month interval. It should also be noted that the Home Assistant platform only sends data points when the measurement changes, for example, if a device sends a measurement with the same value as sent the previous time the value is not sent to the database.
7. *Avg. Data Points per Hour and Device*: the average number of data points recorded per hour and device, this is just the result of dividing the corresponding *Avg. Data Points per Hour* data by the *No. of Devices* data.

Finally, in the last row of Table 6 we show the totals, i.e., the sum, for the data of *No. of Devices*, *Param. per Device*, *Total Param.*, and *Avg. Data Points per Hour*. The total data for *Avg. Data Points per Hour and Device* is omitted since it is not meaningful. The data shows that the total average number of data points received per hour in the system is approximately 25 000. This may seem like a large amount of data, but by using a time-series database the data are handled efficiently.

Time-series databases are optimized for storing and querying time-stamped data, which is common in IoT systems. With this type of database, we are able to efficiently store and retrieve the data in the system, and as a result, the database size is only about 450 MB after more than 8 months of continuous operation of the system. If we compare this to the Home Assistant database, we can see why a dedicated time-series database is needed. As mentioned in the Data storage, processing and visualization Item in Section 2, the Home Assistant uses an SQL database that is not optimized to handle time series data, and as a result the size of our Home Assistant database, which we configured to only store data over a two-week period, is 3.3 GB. Thus, by using a time-series database, our system is able to efficiently handle and store all the data generated over long periods of time.

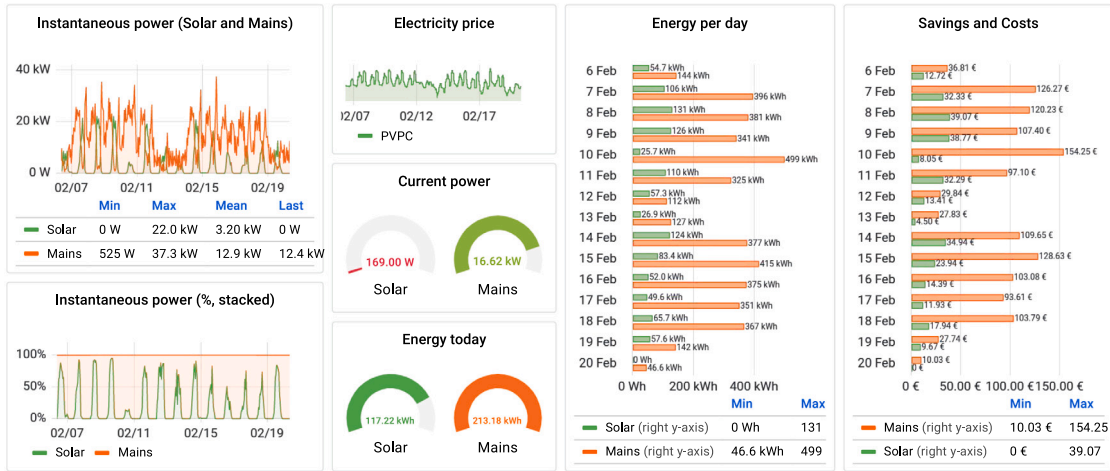

**Fig. 14.** Exemplary Grafana dashboard showing real-time metrics of the solar power system of the Sports Pavilion building.

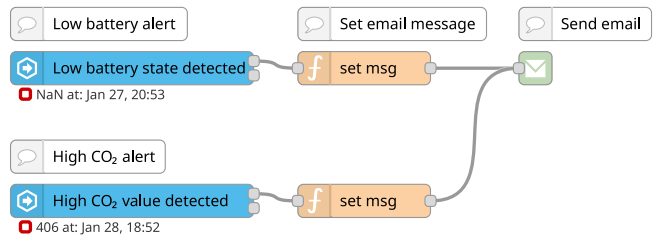

**Fig. 15.** Node-RED flow for sending low battery and high $CO_2$ level alerts.

## 6. Monitoring with Grafana and Node-RED

As stated in Data storage, processing and visualization Item in Section 2, we deployed an instance of Grafana for data visualization purposes. Grafana is an open-source platform for monitoring and analyzing time-series data [21]. It allows users to create interactive and customizable dashboards to visualize data from various sources. In our case, we created several dashboards for each of the integrations of our system according to the needs of the corresponding users, i.e., the occupational safety and health department for the integration in Section 4.1, the CITIC research center for the integration described in Section 4.2, and the architecture, urban planning and equipment department for Sections 4.3 and 4.4. As an example, in Fig. 14 we show one of the Grafana dashboards created for the architecture, urban planning and equipment department to monitor in real-time the power produced by the Sports Pavilion energy system described in Section 4.4.1. This dashboard provides the solar power produced by inverters, the building's power consumption from the electrical grid, daily energy consumption from both sources, and the resulting energy savings. These metrics offer a view of the facility's energy performance, facilitating informed decisions on energy efficiency and sustainability.

Another tool that we used for monitoring purposes was Node-RED. Node-RED is a flow-based programming tool that allows developers to easily create automations and simple applications [22]. It is built on top of the Node.js environment and provides a visual editor for wiring together different nodes, each of which represents a specific function or operation. In the Home Assistant platform, Node-RED is available as a community add-on [56] which can be readily installed. In our system, we used Node-RED in several of the integrations described above to send periodical reports or to send alerts in response to certain events, such as a value exceeding a certain threshold. As an example, in Fig. 15 we show a Node-RED flow used for the integration described in Section 4.2. In this flow, the two leftmost blue nodes are used respectively to detect low battery and high $CO_2$ levels from the connected sensors, if one of such conditions is detected, a message is sent to the corresponding `set msg` node. The `set msg` node will then generate the appropriate alert message and pass it to the rightmost node, which will then send the message to the user via e-mail.

In addition, the architecture, urban planning and equipment department wanted to use the energy data for educational purposes. So, as shown in Fig. 16, displays were installed to show data and information about the generation and consumption of electricity in various buildings. Each display was connected to a small computer and automatically displayed an HTML presentation in a web browser in full screen mode. Such HTML presentations were developed using the reveal.js framework [57] and contain embedded Grafana plots to display real-time data of the monitored systems.

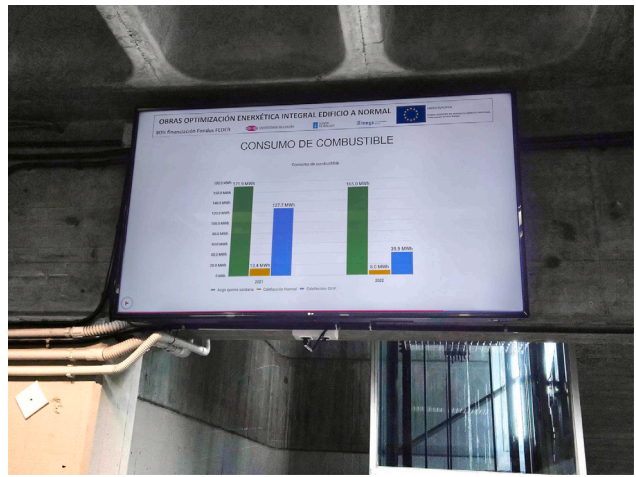

**Fig. 16.** Information displays for educational purposes.

## 7. Published source code

As a further result of this work, the source code of the software developed for the aforementioned integrations has been published in several public repositories. In this way, anyone can make use of this readily available code to integrate one of the scenarios previously described into their IoT systems. The codes developed are the following.

1. We developed the `aranet-cloud` Python module to retrieve data of the Aranet4 sensors from the Aranet Cloud. The module allows to retrieve the most recent data of all the Aranet4 from the Aranet Cloud and also the data over a certain time interval for a given sensor. This code is available in the repository at https://github.com/GTEC-UDC/aranet-cloud-python.
2. We developed the `aranet_to_influxdb` Python module to transfer the Aranet4 sensors data from the Aranet Cloud to an InfluxDB database. This code is available in the repository at https://github.com/GTEC-UDC/aranet-to-influxdb-python.
3. To integrate the BACnet devices described in Section 4.3 we used the `bacnet-stack` library [46]. For this, we created a fork of the library and implemented two new command-line applications. These are:

   - `readpropmjson`: A command-line application based on the `bacnet-stack bacrpm` application. The `bacrpm` application allows to read multiple BACnet properties from BACnet devices. We modified `bacrpm` to return the read values in a JSON format and also added support for connecting to a BACnet controller using the Internet Protocol (IP), which `bacrpm` lacked.
   - `readpropmbyname`: A command-line application based on the previous `readpropmjson` application. To read BACnet properties with `readpropmjson`, the type, object instance number, and numeric identifier of each property must be indicated. The `readpropmbyname` application simplifies this and only requires the property names.

   These codes are available in the repository at https://github.com/GTEC-UDC/bacnet-stack.
4. To retrieve the properties of BACnet objects from a BACnet controller, such as device name, device ID, name, type, units, etc., we developed a simple Python script using the `BAC0` Python library. The information retrieved with this script can then be used with the `readpropmbyname` application described above. This code is available in the repository at https://github.com/GTEC-UDC/bacnet_get_objects_props.
5. We developed the `goodwe-sec1000-info` Python module to retrieve the inverters and electrical grid power data from the GoodWe SEC1000 device described in Section 4.4.1. This code is available in the repository at https://github.com/GTEC-UDC/goodwe-sec1000-info.
6. For the Modbus/TCP integrations described in Section 4.4 we used the `modbus_client` Python library [58], which allows to easily recover typed data from Modbus/TCP devices. We made a fork of the library with some custom modifications and enhancements. This code is available at the repository https://github.com/GTEC-UDC/modbus_client.
7. For the RadonEye Plus2 (RD200P2) sensor described in Section 4.3 we forked the ESPHome [49] project to add support for the sensor. We also submitted a request to the original ESPHome project to integrate our code. Our fork of ESPHome is available in the repository at https://github.com/GTEC-UDC/esphome.

## 8. Conclusion

In this article, we show how we developed an IoT system to monitor and control various heterogeneous devices and systems from multiple vendors. Throughout the design and deployment phases, we kept in mind the key challenges faced in IoT projects, such as interoperability and integration, scalability, and data storage, processing, and visualization. Although these challenges have been previously analyzed in the literature from a theoretical point of view, in this article, we have explained how they influenced the design of our system and how we addressed them in our deployment. This approach allowed us to design an architecture that effectively addresses these challenges and ensures a smooth deployment process.

After deploying the base components of the IoT system, we integrated several devices and systems. At this point, we encountered several interoperability and integration challenges, such as sensor data not being directly available, the need for software licenses, the need to contact customer support, the additional considerations needed when monitoring of wireless devices, protocols not readily available in the deployed IoT platform, proprietary and undocumented protocols, protocol implementations with some limitations or lacking certain features, and non-standard or faulty protocol implementations.

Next, we detailed the specific sensors and systems integrated into our platform. More specifically, we presented five scenarios: The monitoring of air quality (in particular $CO_2$) in our university as a response to reduce the risk of contagion during the SARS-CoV-2 pandemic; the monitoring of several environmental parameters in the building of the CITIC research center; the use of energy efficiency data, distributed among several BACnet controllers that were handled separately in our university; the monitoring of two photovoltaic solar energy systems; and the monitoring of several power meters to measure the university's energy consumption.

In IoT systems, the large volume of data generated by the increasing number of devices can quickly become overwhelming, so it is essential to ensure that resources are appropriately scaled. To check that our system was capable of handling all the data, we analyzed the data and CPU usage statistics and found that the available resources in our system were more than sufficient. This allowed us to ensure that our system was working well and could handle the increasing volume of data as we continue to integrate more devices into our system.

To visualize the collected data, we deployed an instance of Grafana, which provided us with an intuitive and user-friendly interface for monitoring and analyzing the data. In addition, we used Node-RED for monitoring purposes, which is a powerful flow-based tool to process and analyze the data. These tools allowed us to effectively monitor and visualize the data generated by the devices, which was critical to ensure the success of the IoT system.

Finally, we would like to emphasize the importance of releasing the source code for the integrations we have done. This allows other people and organizations to benefit from our work and build upon it.

In conclusion, this work highlights the potential and feasibility of IoT systems in various real-world applications and demonstrates the importance of considering the key challenges faced in IoT projects during the design and deployment phases.

**CRediT authorship contribution statement**

**Tomás Domínguez-Bolaño:** Conceptualization, Software, Visualization, Data Curation, Writing – Original Draft, Writing – Review & Editing. **Valentín Barral:** Software, Writing – Original Draft, Writing – Review & Editing. **Carlos J. Escudero:** Conceptualization, Resources, Writing – Review & Editing, Supervision, Project administration, Funding acquisition. **José A. García-Naya:** Writing – Review & Editing, Supervision, Funding acquisition.

**Declaration of competing interest**

The authors declare that they have no known competing financial interests or personal relationships that could have appeared to influence the work reported in this paper.

**Data availability**

The source code of the software developed during this work is openly available as detailed in Section 7. Other data may be made available on request.

**Acknowledgments**

This work was supported in part by grants PID2022-137099NB-C42 (MADDIE) and TED2021-130240B-I00 (IVRY) funded by MCIN/AEI/10.13039/501100011033; and in part by the European Union NextGenerationEU/PRTR. Funding for open access charge: Universidade da Coruña/CISUG.